\documentclass{aa}
\usepackage{graphicx}
\usepackage{natbib}
\bibpunct{(}{)}{;}{a}{}{,} 
\begin{document}

\title{Extrasolar planet detection by binary stellar eclipse timing: evidence for a third body around CM Draconis}
\author
{H. J. Deeg \inst{1}, B. Oca\~na\inst{1,2}, V. P. Kozhevnikov\inst{3}, D. Charbonneau\inst{4}, F. T. O'Donovan\inst{5}, \and L.R. Doyle\inst{6} }

\offprints
{H.J. Deeg}

\institute{Instituto de Astrof\'\i sica de Canarias, C. Via Lactea S/N, 38205 La Laguna, Tenerife, Spain
\and
Instituto de Radio Astronom\'\i a Milim\'etrica (IRAM), Av. Divina Pastora
7,  N\'ucleo Central, 18012 Granada, Spain
\and
Astronomical Observatory, Ural State University, Lenin ave. 51, 
Ekaterinburg, 620083, Russia
\and
Harvard-Smithsonian Center for Astrophysics, 60 Garden St., Cambridge, MA 02138, USA
\and
California Institute of Technology, 1200 E. California Blvd., Pasadena, CA 91125, USA
\and
SETI Institute, 515 N. Whisman Road, Mountain View, CA 94043, USA\\
\email {hdeeg@iac.es}
}

\date
{Received 6 Nov. 2007; accepted 28 Dec. 2007}

\abstract {New eclipse minimum timings of the M4.5/M4.5 binary CM Dra were obtained  between the years 2000 and 2007. In combination with published timings going back to 1977, a clear non-linearity in observed-minus-calculated (O-C) times has become apparent. Several models are applied to explain the observed timing behavior.}
{Revealing the processes that cause the observed O-C behavior, and testing the evidence for a third body around the CM Dra system.}
{The O-C times of the system were fitted against several functions, representing different physical origins of the timing variations.}
{An analysis using model-selection statistics gives about equal weight to a parabolic and to a sinusoidal fitting function. Attraction from a third body, either at large distance in a quasi-constant constellation across the years of observations or from a body on a shorter orbit generating periodicities in O-C times is the most likely source of the observed O-C times. The white dwarf GJ 630.1B, a proper motion companion of CM Dra, can however be rejected as the responsible third body. Also, no further evidence of the short-periodic planet candidate described by Deeg et al. (2000) is found, whereas other mechanisms, such as period changes from stellar winds or Applegate's mechanism can be rejected.}
{A third body, being either a few-Jupiter-mass object with a period of 18.5$\pm$4.5 years or an object in the mass range of $1.5M_\mathrm{jup}$ to $0.1M_{\sun}$ with periods of hundreds to thousands of years is the most likely origin of the observed minimum timing behavior.} 

\authorrunning{Deeg et al.}

\titlerunning{Extrasolar Planet Detection by Binary Stellar Eclipse Timing}

\maketitle

\section{Introduction}
\object{CM Dra} (\object{LP 101.15}, \object{G225-067}, \object{GJ 630.1}) is a  detached spectroscopic eclipsing M4.5/M4.5 binary with one of the lowest known total masses, of $0.44 M_{\sun}$. With its nearly edge-on inclination of 89.59$\degr$(see \citealt{koz+04} for the most recent orbital and physical elements) it was chosen as the target of the first photometric search for planetary transits. Performed by the 'TEP' project, with an intense observing campaign during the years 1994 - 1999 \citep{ddk+98, ddk+2000}, over 1000 hours of coverage of that system were obtained with several 1m-class telescopes. This lightcurve was initially searched for the presence of transits from planets in circumbinary 'P-type' orbits with 5 - 60 day periods, with a negative result \citep{ddk+2000}. The same lightcurve provided, however, a further possibility to detect the presence of third bodies, from their possible light-time effects on the binary's eclipse minimum times. 

To date, no unambiguously circumbinary planets have been detected\footnote{A possible circumbinary planet is \object{HD202206c}, orbiting around HD202206a and b. This depends however on the classification of the 17.4M$_\mathrm{jup}$ object \object{HD202206b} as being a planet or a brown dwarf \citep{correia+05}}, and their discovery would constitute  a new class of planets. Motivation for this work also arises from previous successes of precise timing measurements to detect the presence of planets. The first known extrasolar planets were detected through light-time effects in the signals of the Pulsar PSR1257+12 \citep{wolyfrail92} and recently, sinusoidal residuals in the pulsation frequency of the sdB pulsating star V391 Peg have been explained through the presence of a giant planet \citep{silvotti+07}, leading to the first detection of a planet orbiting a post-red-Giant star. In both cases, timing measurements have led to detections of planets that would have been difficult or impossible to find with other planet-detection methods, a situation that is similar to the detection of planets around eclipsing binaries -- unless they exhibit transits.

A first analysis of 41 eclipse minima times for CM Dra, presented in \citet{deeg+00} (hereafter DDK00) gave a low-confidence indication for the presence of a planet of 1.5 - 3 Jupiter masses, with a period of 750 - 1050 days. This result was based on a power-spectral analysis of the minimum timings' residuals against a linear ephemeris, indicating a periodic signal with an amplitude of about 3 seconds. Motivated by the result of DDK00, we continued surveying the CM Dra system with occasional eclipse observations in the following years. During this time, it became increasingly clear that a simple linear ephemeris would not provide a sufficient description of the general trend of the eclipse times any longer. This led to the objective of this paper -- a thorough and systematic discussion of the possible processes acting on this interesting system, and an evaluation of the presence of a third body of planetary mass or heavier.

In the following sections, we present the data for this analysis (Sect.~\ref{sec:data}), evaluate the effect of several physical processes on the eclipse timings (Sect.~\ref{sec:processes}), and compare the significance of several numerical fits for the explanation of the observed trend (Sect.~\ref{sec:stats}). This is followed in Sect.~\ref{sec:disc} by a discussion of the physical implications of these fits, with conclusions in Sect.~\ref{sec:conclusion}.

\section{The observational data\label{sec:data}}
The minimum times analyzed here include all minimum timings derived in DDK00. Since its publication, we performed dedicated eclipse observations with the IAC80 (0.8m) and the INT 2.5m telescopes within the Canary Islands' Observatories and with the Kourovka 0.7m telescope of the Ural State University in Ekaterinburg, Russia. In summer 2004, one of the fields surveyed for planetary transits by the TrES collaboration \citep{alonso+07} contained CM Dra and time series spanning several weeks were obtained from the Sleuth 10 cm telescope \citep{sleuth03}. In all cases, photometric time series were obtained from which the eclipse times were measured using the method of \citet{kvw56}. We accepted only results where the formal measurements error from that algorithm was less than 10 seconds, which required data with a largely complete and uninterrupted coverage of individual eclipses. The timings obtained were then corrected to solar-system barycentric (BJD) times with the BARYCEN routine in the 'aitlib' IDL library \footnote{http://astro.uni-tuebingen.de/software/idl/aitlib/} of the University of T\"{u}bingen.

Only a few timings of CM Dra of comparable quality could be found in the literature: \citet{lacy77} gives a total of 9 minimum times from observations in 1976. These timings scatter in O-C by about 30 seconds, and were based on photoelectric data with frequent gaps, including several incomplete eclipses. We therefore chose to remeasure them from \citeauthor{lacy77}'s tabulated lightcurve with the same procedures used for our own data, from which only two timings of sufficient quality could be obtained for the further analysis. Only two additional timings from the years 2004 and 2005 could be found in the literature \citep{IBVS5745, IBVS5603}, both in good agreement with the other data. The final data set contains 63 minima, of which 27 are primary and 36 are secondary eclipses. \\

Since these minimum timings cover about 30 years of observations and are of consistently high precision, the effects from the 18 leap-seconds that have been introduced into Universal Time during that span need to be corrected for. Consequently, all minimum times used in this work were converted from the conventional UT to TAI (International Atomic Time)\footnote{The conversion TAI - UTC can be obtained from ftp://maia.usno.navy.mil/ser7/tai-utc.dat} which is a timescale with constant and uniform flow, without discontinuities from leap-seconds. All these minimum times are listed in  Table~\ref{tab:mintimes} together with the formal errors of the minimum times, the cycle number E and O-C (observed - calculated) residuals against the ephemerides of \citet{ddk+98}. For data newly presented in this work, the originating telescope is also indicated.\\

\begin{table*}
\caption{CM Dra eclipse minimum times, given in International Atomic Time (TAI)}             
\label{tab:mintimes}      
\centering          
\begin{tabular}{l c c c c c l }     
\hline\hline       
$T_\mathrm{min}$\,(TAI)&$\sigma_{T\mathrm{min}}$&Prim/Sec&E&O-C&Source&Instrument\\
BJD-2400000&($10^{-6}$\,d)&&&(sec)\\
\hline                    
 42893.932818&51&I&-5469&-31.8&1&-\\
42994.768729&36&II&-5390&-24.7&1&-\\
49494.634037&32&I&-265&1.2&2&-\\
49497.803721&55&II&-263&-9.6&2&-\\
49499.707562&28&I&-261&-1.7&2&-\\
49500.975836&73&I&-260&-11.7&2&-\\
49501.609080&103&II&-260&6.7&2&-\\
49511.756133&58&II&-252&1.1&2&-\\
49562.491845&27&II&-212&11.3&2&-\\
49815.536668&87&I&-12&1.0&2&-\\
49828.853584&24&II&-2&-0.2&2&-\\
49830.757404&14&I&0&5.9&2&-\\
49833.927084&17&II&2&-5.3&2&-\\
49840.904433&35&I&8&-1.9&2&-\\
49853.588403&47&I&18&4.3&2&-\\
49855.489863&33&II&19&7.7&2&-\\
49858.661903&38&I&22&-0.9&2&-\\
49872.614147&44&I&33&-4.6&2&-\\
49881.492866&47&I&40&-5.5&2&-\\
49947.449171&18&I&92&-2.8&2&-\\
49949.350561&15&II&93&-5.4&2&-\\
49954.424251&33&II&97&5.9&2&-\\
50221.421524&16&I&308&9.6&2&-\\
50222.689653&27&I&309&-12.9&2&-\\
50243.617062&102&II&325&-0.7&2&-\\
50244.885447&53&II&326&-1.1&2&-\\
50252.495738&32&II&332&-5.3&2&-\\
50257.569367&36&II&336&0.7&2&-\\
50259.473057&51&I&338&-4.4&2&-\\
50262.642977&33&II&340&5.1&2&-\\
50272.790006&31&II&348&-2.7&2&-\\
50636.818094&41&II&635&14.4&2&-\\
50643.795304&35&I&641&5.8&2&-\\
50650.770213&55&II&646&-0.3&2&-\\
50710.384658&40&II&693&10.2&2&-\\
50905.716634&21&II&847&4.8&2&-\\
50993.870837&61&I&917&-0.1&2&-\\
50995.772307&48&II&918&4.2&2&-\\
51000.845877&42&II&922&5.1&2&-\\
51009.724607&49&II&929&5.2&2&-\\
51340.774332&28&II&1190&2.8&2&-\\
51359.800251&26&II&1205&8.9&2&-\\
51373.752531&18&II&1216&8.2&2&-\\
51616.650417&37&I&1408&13.6&3&IAC-80\\
51766.320420&42&I&1526&13.6&3&Kourovka\\
51780.272729&30&I&1537&15.3&3&Kourovka\\
52416.369110&42&II&2038&18.0&3&Kourovka\\
52799.422892&15&II&2340&21.8&3&INT\\
52853.330664&29&I&2383&25.1&3&Kourovka\\
53082.909268&100&I&2564&28.4&4&-\\
53117.788867&89&II&2591&32.1&3&Sleuth\\
53120.960987&11&I&2594&30.4&3&Sleuth\\
53136.814607&46&II&2606&22.7&3&Sleuth\\
53138.718577&66&I&2608&41.8&3&Sleuth\\
53145.693367&74&II&2613&25.4&3&Sleuth\\
53153.939177&73&I&2620&35.0&3&Sleuth\\
53160.914217&97&II&2625&40.3&3&Sleuth\\
53456.448870&55&II&2858&24.4&3&Kourovka\\
53472.305006&34&I&2871&32.8&3&Kourovka\\
53478.647065&100&I&2876&42.3&5&-\\
53498.305789&62&II&2891&29.0&3&Kourovka\\
53503.379347&42&II&2895&28.9&3&Kourovka\\
54166.747285&37&II&3418&32.3&3&IAC-80\\
\hline  
\end{tabular}
\begin{list}{}{}
\item Sources: 1: Remeasured from lightcurve of \citet{lacy77}, 2: \citet{deeg+00}, 3: This work, 4: \citet{IBVS5603}, 5: \citet{IBVS5745}.
\end{list}
\end{table*}

  \begin{figure}
   \centering
    \includegraphics[width=9.0cm]{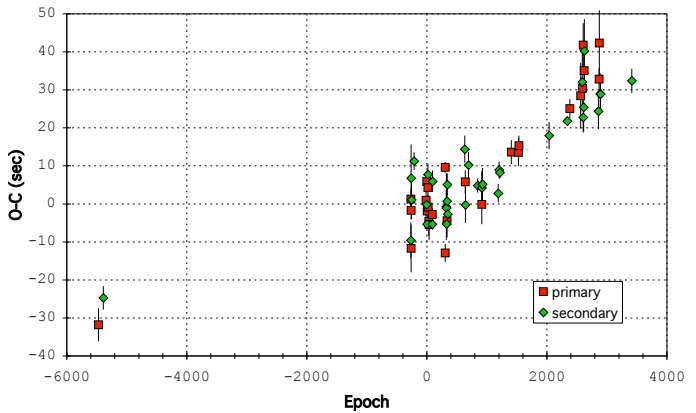}
  
   \caption{O-C values of CM Dra eclipse minimum times from Table~\ref{tab:mintimes}, against ephemerides from DDK00}
              \label{fig:OCtimes}
    \end{figure}

\section{Physical processes' effects on O-C times\label{sec:processes}}
The further analysis of the eclipse minimum times is based on the analysis of the temporal development of the 'O-C' residuals between observed and calculated (expected from ephemeris) eclipse minimum times. In these, 
\begin{equation}
(O-C)_E = T_E - T_{c,E}\ ,
\end{equation}
where $T_{c,E}$ refers to a minimum time calculated from an ephemerides at 'Epoch' or cycle number $E$, and $T_E$ refers to the corresponding observed minimum time. The observed times $T_E$ are related to the minimum times $T'_E$ in the  binary system's rest frame by $T_E = T'_E + d_E/c$, where $d_E$ is the distance from the observer to the binary at cycle $E$ and $c$ is the velocity of light. Hence, based on  a linear ephemeris 
$T_{c,E} = E P_c + T_{c,0}$, where $P_c$ and $T_{c,0}$ are the ephemerides' period and time of conjunction. The difference, $(O-C)_E$, is given by: 
\begin{eqnarray}
(O-C)_E &=&T'_E + d_E/c - E P_c - T_{c,0}
\label{eq:basicOC}
\end{eqnarray}
This general expression allows us now to develop the cases to be considered in our analysis. 
 
\subsection{Accelerated binary systems with constant period\label{sec:accel}}

First, we review the case of a binary system moving at a constant velocity $v_0$ relative to the observer, with $v_0 << c$. The distance to this system is given by $d_E = d_0 + v_0 E P'$, and the minimum times in the binary system's rest-frame are given by $T'_E = T'_0 + E P'$, where $P'$ is constant. The sub-indices '$0$' refer to the parameters' values at the moment when $E=0$. From Eq.~\ref{eq:basicOC}, we obtain then a relation linear in $E$:
\begin{eqnarray}
(O-C)_E &=& T'_0 + E P' + (d_0 + v_0 E P')/c - E P_c - T_{c,0} \\
&= & (T_0 - T_{c,0}) + E(P - P_c) 
\label{OCconstvel}
\end{eqnarray}
where $P = P'(1+v_0/c)$ is the observable period and $T_0 = T'_0 + d_0/c$ is the minimum time that  should have been observed at $E=0$. Hence, in a system where the observed O-C times deviate linearly from the ephemeris, the terms $\kappa_0= (T_0 - T_{c,0})$ and $\kappa_1=(P - P_c)$ indicate errors in the derivation of the original ephemerides $(T_{c,0},P_c)$, but do not have any further physical meaning.\\

If we consider as $D(E)$ any additional distance to the binary that cannot be expressed by a linear term, so that $d_E = d_0 + v_0 E P' + D(E)$, then the O-C times will be modified by the 'light-time effect' $D(E)/c$:
\begin{eqnarray}
(O-C)_E = & (T_0 - T_{c,0}) + E (P - P_c)  + D(E)/c\ .
\label{eq:OCwithD}
\end{eqnarray}
Hence, the non-linear components of $(O-C)_E$ describe the non-linear components of the distance to the binary. The non-linear distance component $D$ has necessarily to be caused by an acceleration process, with $D(E) = \int\int_{T'_0}^{T'_0 + E P'} a_\parallel(t)\  dt dt$, where $a_\parallel$ is the acceleration along the line of sight to the binary. Within the scope of this work, the following acceleration scenarios have been considered:\\

\emph{i}) For the case of a constant acceleration $a_\parallel$, with $D(E) =\frac{1}{2}a_\parallel (EP')^2$  we obtain now:
\begin{eqnarray}
(O-C)_E &= & (T_0 - T_{c,0}) + E (P - P_c) + E^2 a_\parallel \frac{P'^2}{2c}\\
	& = &\kappa_0 + E\kappa_1 + E^2\kappa_2
	\label{eq:OCconstacc}
\end{eqnarray}
where the $\kappa_i$ refer to the polynomial coefficients of the fit given in Table~\ref{tab:fitres}. \\

\emph{ii}) In the case of an acceleration that undergoes a constant change $\dot{a}_\parallel$, the O-C times behave like a third order polynomial:
\begin{eqnarray}
(O-C)_E 	& = &\kappa_0 + E\kappa_1 + E^2\kappa_2 + E^3\kappa_3\ 
\end{eqnarray}
where  $\kappa_3$ is given by:
\begin{eqnarray}
\kappa_3= \dot{a}_\parallel  \frac{P'^3}{6c} \ 
\end{eqnarray}
whereas the coefficients $\kappa_0 - \kappa_2$ remain identical to the previous case.

\emph{iii}) For a binary that is accelerated due to third body on a circular orbit, the amplitude of the timing variation $D/c$ is then given by:
 \begin{equation}
D/c = \frac{m_3\ d_{\parallel b3}}{(m_b + m_3)\ c}=\frac{m_3}{(m_b+m_3)^{2/3}c} \left( \frac {P_3^2 G}{4\pi^2} \right)^{1/3} \sin i
\label{eq:sinampli}
\end{equation}
where $d_{\parallel b3}$ is the line-of-sight-component of the distance between the barycenter of the binary and the third body, and $m_3$ and $m_b$ are the masses of the third body and the binary stars, respectively. In the right hand term, $d_{\parallel b3}$ has been substituted for $P_3$, the period of the third body, with $i$ being the inclination of its orbital plane and $G$ is the gravitational constant. A development of  the general case is given by \citet{irwin59}, with further examples of recent applications in \citet{demircan02}.

From Eq.~\ref{eq:sinampli}, the O-C times are then given by
\begin{eqnarray}
(O-C)_E & = &  \kappa_0 + E\kappa_1 + \kappa_d \cos ( E \kappa_\phi+  \phi_0)
 \label{eq:ocsine}
 \end{eqnarray}
with $\kappa_d =D/c$, and the term $E \kappa_\phi+  \phi_0$ describing the phase of the third body, where 
\begin{equation}
\kappa_\phi =  \frac{P' 2 \pi}{P_3}
\label{eq:kappaphi}
 \end{equation}
 and $\phi_0$ is the phase of the third body at $T_0$.

\subsection{Variation of the intrinsic binary period\label{sec:varper}}
Here we consider the consequences of a period variation of a binary moving at a constant velocity. The intrinsic period variation $\partial P^{'}/\partial E$ shall be small enough to be considered constant across the observed time span. The period at cycle $E$ is then given by  $P^{'}_E = \int \frac{\partial P^{'}}{\partial E} \,dE + P^{'} _{0} = E \frac{\partial P^{'}}{\partial E} + P^{'} _{0}$, and the times of minima in the binary's reference frame are:
\begin{eqnarray}
T^{'}_{E}& = &\int P^{'}_E\,dE + T^{'} _{0}\\
& = &\frac{1}{2}E^2  \frac{\partial P^{'}}{\partial E} + E  P^{'} _{0} + T^{'} _{0}
\end{eqnarray}
After converting from the times $T^{'}_{E}$ to the observable times $T_E$ by inserting this equation into Eq.~\ref{eq:basicOC} , we obtain a quadratic equation that is similar to the accelerated system described by Eq.~\ref{eq:OCconstacc}, with the only difference being that the parameter $\kappa_2$ is now given by:
\begin{equation}
\kappa_2= \frac{1}{2}  \frac{\partial P^{'}}{\partial E}(1+\frac{v_0}{c}) \ .
\label{eq:periodvar}
\end{equation}

\section{The observed minimum times \label{sec:stats}}

As in DDK00, O-C times were derived using the linear ephemerides given by \citet{ddk+98}, which was based on a fit to eclipse timings observed from 1994 to 1996.  For the present work, this ephemerides was converted to TAI by adding 29 seconds, which corresponds to the difference TAI-UT that was in effect during most of that time (July 1, 1994 to Dec 31, 1995). The O-C values of DDK00 differ therefore by a maximum of only 1 sec between the original work in DDK00 and the present work. The ephemerides conversion from \citet{ddk+98}  to TAI is:
\begin{eqnarray}
T_{c\mathrm{I}} &= &2\,449\,830.75734 \pm 0.000\,01 + P_c \ E\;(TAI)\\
T_{c\mathrm{II}} &= &2\,449\,831.39037 \pm 0.000\,01 + P_c \ E\;(TAI)
\end{eqnarray}
where $T_{c\mathrm{I}}$ and $T_{c\mathrm{II}}$ refer to primary and secondary minima, respectively, $E$ is the epoch or cycle number, and the period is given by $P_c = 1.268\,389\,861 \pm 0.000\,000\,005\ $days. The O-C values against that ephemerides are given in Table~\ref{tab:mintimes} and shown in Fig~\ref{fig:OCtimes}. Since the writing of DDK00, a clear trend of increasing O-C values has become apparent, which is also apparent from the extension into the past through the inclusion of the values from \citet{lacy77}, which weren't included in DDK00's analysis. The linear ephemerides of DDK00, therefore, no  longer provides the best description of the observed O-C values. However, we chose to maintain this ephemerides, since small errors in the parameters of a linear ephemerides have no physical meaning in the interpretation of higher-order O-C dependencies, as was shown in Eq.~\ref{OCconstvel}.

 \subsection{Temporal evolution of the phase of the secondary eclipse}
A primary concern was assuring that the primary and secondary eclipses were not undergoing any evolution of their relative phase due to, for example, variations in eccentricity or in the argument of periastron due to apsidal motion. Therefore we investigated if the orbital phase of the secondary eclipse underwent any variations. This was done through a comparison of two sub-samples of the timings from Table~\ref{tab:mintimes}, taking an early one from Epochs 0 to 400 and a late one from Epochs 2500 - 2900. In both samples, the average of the primary eclipses was set to zero, and the corresponding phase of the secondary eclipses were calculated, giving: 

\begin{equation}
\mathrm{phase }=0.499064  \pm 0.000017 \mathrm{\ for\ E=0 - 400}
 \end{equation}
\begin{equation}
\mathrm{phase }=0.499039 \pm 0.000029 \mathrm{\ for\ E= 2500 - 2900\ .}
\end{equation}
The difference between these two values is well within their error-bars; hence no relevant change in the phase of the secondary eclipse was detected. Since primary and secondary eclipses of CM Dra have very similar depths, resulting in timing measurements of similar precision, both types of eclipses were treated together and equally in the further analysis.

\begin{table*}[!ht]
\begin{center}
\caption{Parameters of fits to O-C times}
\label{tab:fitres}
\scriptsize{
\begin{tabular}{l c c c c c c c c c c}
\noalign{\smallskip}
\hline\hline
\noalign{\smallskip}
\textbf{ }&$\kappa_0$&$\kappa_1$&$\kappa_2$&$\kappa_3$&$\kappa_d$&$\kappa_\phi$&$\phi_0$&$\chi^2$&AIC$_c$&$w_i$\\
&s&$10^{-3}$s&$10^{-7}$s&$10^{-10}$s&s& $10^{-3}$rad & rad&-&-&-\\
\noalign{\smallskip}
\hline\\
Linear &       $3.6\pm 1.0$  & $7.3\pm0.4$ &-&-&-&-&-&37.94&49.4&0.002\\
\multicolumn{7}{l}{$O-C=\kappa_0+E\kappa_1$}\\ \hline \\
Parabola   & $-1.5\pm 1.5$& $8.7 \pm 0.5$  &$7.0 \pm 1.3$&-&-&-&-&11.57&37.4&0.649\\
\multicolumn{7}{l}{$O-C=\kappa_0+E\kappa_1+E^{2}\kappa_2$} \\ 
\hline \\
Third Order Polynomial   & $-1.1\pm 1.7$& $8.0 \pm 2.0$   &$7.8 \pm 2.6$&$0.4\pm 1.1$&-&-&-&11.45&41.6&0.079\\
\multicolumn{7}{l}{$O-C=\kappa_0+E\kappa_1+E^{2}\kappa_2+E^3\kappa_2 $} \\ 
\hline \\
Linear + Sine&$7.3_{-4.2}^{+2.1}$ & $5.3_{-0.9}^{+2.1}$ &-&-& $8.7_{-3.0}^{+2.2}$&$1.18_{-0.24}^{+0.36}$&$4.1_{-0.7}^{+0.4}$ &6.18&39.1&0.270 \\
\multicolumn{1}{l}{$O-C=\kappa_0+E\kappa_1+\kappa_d \sin(E \kappa_\phi+\phi_0)$}\\
\noalign{\smallskip}
\hline
\hline
\end{tabular}
}
\normalsize
\rm
\end{center}
\end{table*}

\subsection{Model fits to the O-C times\label{sec:modelfit}}
Our initial intent was a direct fitting of the functions described in Section~\ref{sec:processes} to the O-C residuals of Table~\ref{tab:mintimes}. However, in the subsequent statistical analysis we noted two problems: First, the average formal error in the individual O-C measurements from Table~\ref{tab:mintimes} is 3.90 seconds.  On the other hand, the standard deviation among the residuals (observations - fits) could not be reduced below 5.6 seconds. Reduced $\chi^2$ values were not less than  $\chi_\mathrm{red}^2 \approx 4.2$, even when applying 5th or 6th order polynomial fits, whereas a 'good' fit should indicate values of $\chi_\mathrm{red}^2 \approx 1$. Since the fits are not intended to  -- and cannot --  model the point-to-point variations and furthermore, since the presence of periodic short-frequent O-C timing variations can be ruled out (see section~\ref{sec:powerspectra}), we concluded that the formal errors are sub-estimating the real measurement errors. In the further analysis, a value of 5.6 seconds was adopted as a minimum measurement error and errors smaller than this were set to 5.6 sec. With fourth to sixth order polynomial fits to these data we now obtain $\chi_\mathrm{red}^2 \approx 1$. 

\begin{figure}
   \centering
   \includegraphics[width=9.0cm]{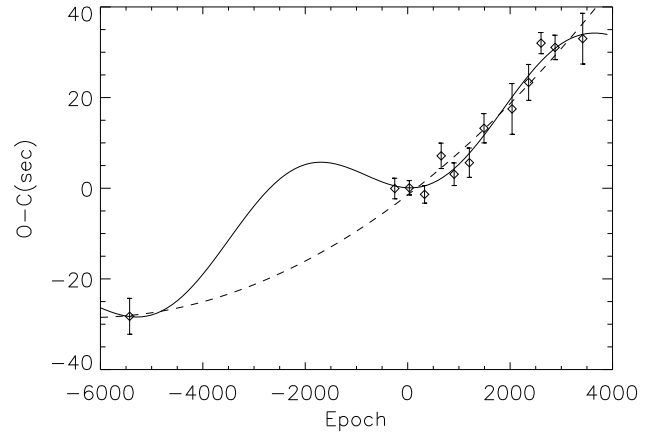}
   \caption{Seasonally averaged O-C values of CM Dra, with parabola fit (dashed line) and sine-linear fit (solid line)}. 
              \label{fig:OCbinfit}
    \end{figure}

A second problem arose because the data-points aren't uniformly or randomly distributed, but clustered into yearly observing  seasons. These clusters act like pivots for the fitting functions, reducing the degrees of freedom of the model fits over the number of data points. The choice of the correct number of degrees of freedom is, however, important for the model comparison performed in Sect.~\ref{modelsel}. Consequently, for each year of observations, we generated one single data point from a weighted average of each season's points. The resulting binned O-C  times are shown in Fig.~\ref{fig:OCbinfit}. 

Fits of the functions that have been discussed in Sect.~\ref{sec:processes} were then performed against this sample, with their best-fit parameters given in Table~\ref{tab:fitres}. The fits were obtained with the IDL library routines "POLY\_FIT" for the polynomial fits, and "AMOEBA" (based on \citealt{press+92}) for the sinusoidal fit, with both algorithms performing minimizations of the $\chi^2$ values. It should be noted that the aim of the sine-linear fit was to test if the data can be well fit to one or a few cycles of such a function, with a period longer than about 10 years. For a search for shorter periodicities see Sect.~\ref{sec:powerspectra}.

The parameters' errors given in Table~\ref{tab:fitres} are 1-sigma confidence limits, whose derivation is described in the next section. Furthermore, the Table's last three columns give the best-fit $\chi^2$ values, calculated for a sample standard deviation of $s=3.2$ s, the Akaike Information Coefficient AIC$_c$, and the Akaike weights $w_i$; both of which will be  introduced in Sect.~\ref{modelsel}.
 
 \subsection{Errors of fit-parameters}
For an estimation of the errors of the fit-parameters we applied initially the common resampling or bootstrap method (e.g. \citealt{cameron+07}), consisting of repeated model fits to a synthetic data set.  It requires one to assume some reference function that describes correctly the general trend of the data, against which residuals of the data points are generated. The synthetic data are generated by permutating the residuals among the data points. The model to be evaluated is then fitted against the synthetic data and distributions of the obtained fit-parameters are used to estimate the likelihood distribution of the parameters from the model-fit on the original data. This method is easy to implement and leads to synthetic samples with properties similar to the original data, but in this work's context two problems arose:\\
First, the analysis is based on the assumption that the distribution of the fit-parameters on the synthetic sets is identical to the likelihood distribution of the parameters obtained from the fit of the original data, which is far from certain.\\ 
Second, resampling is based on the assumption that the reference function is a correct description of the trend of the data, and that residuals against it are measurement errors. This approach may be justified if the underlying physical model - and the function that describes it - is known, and only a refining of parameters is required. In this work however, we are also faced with an uncertainty about the nature of the reference function.  

A method that overcomes these problems, and which has come to the awareness of astrophysicists in recent years, is the  Markov Chain Monte-Carlo (MCMC) method (e.g. \citealt{tegmark+04,ford05,holman+06,burke+07}). MCMC is based on the states of a Markov Chain undergoing random variations, whose probabilities are however directed by the maximum likelihood estimator (MLE) of the fit-parameters (relative to the original data) at each step in the chain. The frequency of states of the Markov Chain at a given parameter value indicates then the posterior probability distribution of that parameter. We refer to \citet{ford05}  for further references to the MCMC, as well as for the implementation of the MCMC with the Metropolis-Hastings Algorithm that was used in our data analysis.

\begin{figure}
   \centering
  \includegraphics[width=9.0cm]{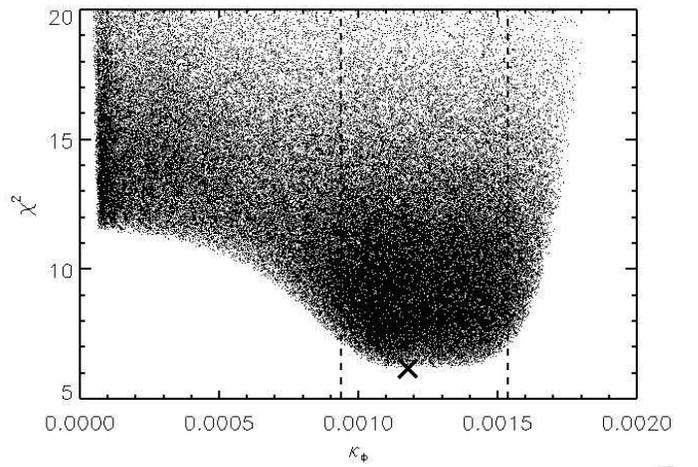}
   \caption{Distribution of the $\kappa_{\phi}$ parameter from the sine-linear model against $\chi^2$ from a Markov Chain of 20 million steps. The dashed lines indicate the parameter's 1-$\sigma$ confidence region, corresponding to $\Delta\chi^2 = 1$ over the best fit's $\chi^2$ value (cross)}. 
              \label{fig:kphi_chi}
    \end{figure}
    
Our final error-analysis is however not based on the posterior probability distributions derived from histograms of the parameter distributions. Instead we have used the MCMC as a tool to explore the relation of $\chi^2$ against the multivariate parameters.  Similar to \citet{burke+07}, for any recorded step of the MCMC chain, the encountered parameters and the corresponding $\chi^2$ values were registered. As an example, Fig.~\ref{fig:kphi_chi} shows the relation between the $\kappa_\phi$ parameter in the sine-linear model fit against $\chi^2$. There is a clear lower limit of $ \chi^2$ for a given parameter value. With increasing numbers of iterations in the Markov Chain, this lower limit approaches the best possible fit at a given parameter value. Hence, the lower limits of $\chi^2$ may be used as tracings of the best-fit $\chi^2$ against one or multiple parameters. Since the distribution-density of a MCMC sequence gravitates towards the regions of lowest $\chi^2$, the MCMC method may therefore be used as a simple tool to trace low-sigma confidence regions around the best-fit parameters. This application of the MCMC also avoids a problem that easily occurs in the interpretation of posterior probabilities from the parameter densities. As can be seen in Fig.~\ref{fig:kphi_chi}, the distribution of points also has dense zones close to the left cutoff, although the minimum value of $\chi^2$ in that zone corresponds to very poor fits. With the flat dependency of the best-fit $\chi^2$ against $\kappa_\phi$ in that zone, the Markov Chains had difficulties  returning to better-fitting values. In the given case, the Markov Chains instead went into 'exploring' a wide multi-parameter space of poor models that opened up close to the left cutoff in Fig.~\ref{fig:kphi_chi}.

For the errors indicated in Table~\ref{tab:fitres} we used chains with a length of 20 million iterations, after discarding the first 1000 steps of the chains. \\

\subsection{Model selection\label{modelsel}}
After applying the fits of several models, physical interpretations will need some information about the likelihood that a given fit corresponds to the true behavior of the observed data. This question is commonly referred to as 'model selection'. Values such as $\chi^2$ (see Table~\ref{tab:fitres}), or the 'reduced Chi-square' of $\chi^2 / \nu$, where $\nu$ is the degrees of freedoms, give a general indication of the quality of a fit. However,  in the case of small or moderate differences of fit-quality among models, they serve little to generate statements that are useful for the model selection.

Probably the best-known method of assigning likelihoods in fit-comparisons is the 'F-test'. It is however valid only  for the comparison of nested models, such as polynomials of different orders, and was therefore not used further. The Akaike Information Coefficient (AIC, \citealt{akaike74}) does not have this limitation, which led to its use in the further investigation. For an introduction to the use of the AIC as a tool for model selection, we refer the reader to \citet{burnham+04,liddle07,mazerolle04}. 
Using the residuals' squared sum (RSS) as the likelihood estimator, where $\mathrm{RSS} = \sum{(y_i - f_i)^2}$ with $y_i$ being the data and $f_i$ the model values, the AIC is given by:
\begin{equation}
\mathrm{AIC} = -n \ln (\mathrm{RSS}/n) + 2 k
\end{equation}
where  $n$ is the number of observations and $k$ is the number of model parameters. Here we use the generally preferred \citep{burnham+04} second order corrected coefficient 'AIC$_c$', which is valid for both small and large samples:
\begin{equation}
\mathrm{AIC}_c = \mathrm{AIC} + \frac{2k(k+1)}{N-k-1}\,.
\end{equation}
While smaller AIC values indicate better fits, their absolute values don't have any meaning.  They are only useful if values from different models are compared. If the best among several models has a value of ${AIC}_{c,\mathrm{min}}$, then for any model $i$ the differences $\Delta_i = \mathrm{AIC}_{c,i} - \mathrm{AIC}_{c,\mathrm{min}}$ may be calculated. Differences  of $\Delta_i \le 2$ indicate substantial support (evidence) for model $i$; models where  $4 \le \Delta_i \le 7$ have considerably less support, while models with $\Delta_i \ge 10$ have essentially no support  
\citep{burnham+04}.  For the comparison within a set of $R$ models, normalized Akaike weights may be derived for each model $i$, with
\begin{equation}
w_i = \frac{\exp(-\Delta_i/2)}{\sum_{r=1}^R \exp (-\Delta_r/2)} \,
\end{equation}
where all weights $w_i$ sum up to 1 (see Table~\ref{tab:fitres}). Following \citet{akaike81}, these weights may be interpreted as a likelihood that can be assigned to each of the models, with the parabolic fit being most likely, followed closely by the sine-linear fit. A word of caution, also reflected in several references about this topic (e.g. \citealt{liddle07}), should however be given against the use of these statistical values as a strong argument in favor of one or the other model: There are several alternative indicators available, such as the Bayesian Information criterium (BIC, \citealt{schwarz78}) or the  Deviance Information Criterion (DIC, \citealt{DIC02}), with different 'penalizations' for models with additional degrees of freedom. While these criteria indicate similar preferences for models with well separated $\chi^2$ or AIC$_c$ values, interchanges in ranking may happen among models that are close in AIC$_c$ values. This cautionary position was backed by a calculation of the BIC for our models. In that case, the sine-linear fit was ranked best, with a slightly lower (and hence 'better') BIC than the parabola fit, while the ranking of the other models remaining unchanged. 

In summary, both the simple linear fit and the third order polynomial fit have significantly less support than the parabolic fit and the sine-linear fit. In Section~\ref{sec:disc} we therefore focus on the physical implications from these two top-ranked models.

\section{Discussion: Possible causes of the observed  O-C times\label{sec:disc}}
Common to all fits, the linear parameter $\kappa_1$ has fairly similar values indicating clearly that the average period of CM Dra across the recorded observations is several milliseconds longer than given by \citet{ddk+98}. 
As shown in Sections~\ref{sec:accel} and \ref{sec:varper}, a parabolic O-C function may arise from two causes, an intrinsic variation in the system's period, or a light-time effect from a \emph{constant} acceleration of the entire binary system. In both cases, the only interesting parameter is the quadratic term, found to be (see Table~\ref{tab:fitres}) $\kappa_2 = (7.0 \pm 1.3) 10^{-7}$ sec/period.

\subsection{Intrinsic period variation}
Considering an intrinsic period variation, the change in period-length per cycle is given by Eq.\ref{eq:periodvar} with $\frac{v_0}{c_l} << 1$ as:
\begin{equation}
 \frac{\partial P^{'}}{\partial E}  \approx 2 \kappa_2 = (1.4\pm0.26)\times10^{-6} \mathrm{s/period},
\end{equation}
The corresponding unitless period change per time is given by:  
\begin{equation}
\frac{\partial P^{'}}{\partial t}=\frac{\partial P^{'}}{\partial E} \frac{1}{P^{'}}=1.28\times10^{-11}
\label{eq:dpdtcmdra}
\end{equation}

\citet{demircan+06} performed a statistical study of the orbital parameters of a sample of detached chromospherically active eclipsing binaries of different ages. Out of that sample, they concluded that their periods \emph{decrease} with an average value of $\alpha=3.96\times10^{-10} yr^{-1}$, with $\alpha$ defined by the differential equation $dP/dt = - \alpha P$. This value may be considered constant throughout a large part of a binary's evolution and is due to angular momentum loss from magnetically driven stellar winds. The corresponding period change of CM Dra would be $\frac{\partial P^{'}}{\partial t}=-1.38\times10^{-12}$, obtained by multiplying $-\alpha$ with CM Dra's period in units of years ($3.47\times10^{-3}$yr). This value is of opposite sign than the observed one (Eq.~\ref{eq:dpdtcmdra}), and is an order of magnitude smaller. Hence, angular momentum loss may well be present, but is not detectable in the current minima timings.
 
 \subsection{Acceleration due to a quasi-stationary attractor}
The second source for a parabolic shape in an O-C diagram could be a constant acceleration of the binary along the line of sight, with changes of acceleration strength and direction during the span of observations being negligible. The acceleration term is given from  Eq.~\ref{eq:OCconstacc} as:
\begin{equation}
a_\parallel = 2 \frac{c \kappa_2}{P^{'2}} = (3.5\pm0.7)\times10^{-8}\mathrm{m/s}^2
\label{eq:accval}
\end{equation} or
\begin{equation}
\frac{a_\parallel}{c}= (1.17\pm0.22)\times10^{-17}\mathrm{s}^{-1}
\end{equation}

We note that this acceleration is at least two orders of magnitude larger than the acceleration of the Solar system, currently constrained within a few$\times 10^{-19}\mathrm{s}^{-1}$ \citep{zakamska05}. 
A quasi-constant acceleration may be caused by a third body of mass $m_3$ at a distance far away enough so that mutual orbital motions don't lead to significant changes in the acceleration vector. The acceleration on the binary caused by a third body is given by:
\begin{equation}
\mathbf{a}= \frac{Gm_3}{r^3} \mathbf{r}
\label{eq:accel}
\end{equation}
where $\mathbf{r}$ is a distance vector from the barycenter of the binary towards the third body. We note that this acceleration is independent of the mass of the binary. The acceleration component along the line of sight is then given by
\begin{equation}
a_\parallel =  \frac{Gm_3}{r_\perp^2}\cos^2 i \sin i
\end{equation}
where $i$ is the inclination, defined here as the angle between  $ \mathbf{r}$ and the plane of the sky, and $r_\perp= r \cos i$ is the lateral component of $r$. The equation above gives $a_\parallel = 0$ for inclinations of both $0\degr$ and $90\degr$ and a maximum for $a_\parallel$ at inclinations of $i = \arctan \sqrt{1/2} = 35.26\degr$, leading to 
\begin{equation}
a_{\parallel} \le 0.3849\  \frac{Gm_3}{r_\perp^2}. 
\end{equation}
Converting to common astronomical units, the minimum mass $m_3$ at a given lateral distance is then obtained by 
\begin{equation}
\left(\frac{m_3}{M_{\sun}}\right) \ge 438.26 \left(\frac{a_\parallel}{m s^{-2}}\right)  \left(\frac{r_\perp}{AU}\right)^2 . 
\label{eq:m3}
\end{equation}

Replacing $r_\perp$ by an angular separation based on the distance to CM Dra (15.93 pc; \citealt{chaYbar95}), and using $a_\parallel =3.5\pm0.7\times10^{-8}\mathrm{m/s}^2$,  we may now derive the minimum mass of a possible third body at a given angular separation from CM Dra:
\begin{equation}
\left(\frac{m_3}{M_{\sun}}\right) \ge 0.0030 \left(\frac{\alpha}{\mathrm{arcsec}}\right)^2 . 
\label{eq:minmass}
\end{equation}

The third order polynomial fit, while resulting in a lower weight in the model selection, is not to be discarded completely. It would describe a system undergoing a constant change in acceleration $\dot{a}_\parallel$. There is however no physical process that generates a truly constantly varying acceleration. Hence the third-order polynomial fit may only describe cases where $\dot{a}_\parallel$ is quasi-constant across the observing time-span. The third order polynomial fit may, however, provide the first terms of a Taylor-expansion of the true acceleration process of the system. For a slowly changing acceleration, like a cyclic one with long periods of O(100 yr) or longer, the 3rd order polynomial may therefore give a better description of the O-C times than the parabolic fit does. From our fit, however, with the value of $\kappa_2$ being very close to the one from the parabolic fit, the derived acceleration $a_{\parallel}$ and the constraint for a third mass given in Eq.~\ref{eq:minmass} do not significantly differ.

A possible source for an acceleration of the CM Dra system may be the nearby white dwarf \object{GJ 630.1B } (\object{WD 1633+57},  \object{LP 101.16},  \object{G225-068}). This has long been recognized as a proper-motion companion to CM Dra \citep{giclas+71} at an angular distance of 26$\arcsec$, which corresponds to a lateral distance $r_\perp$ of 414AU. With a period of $O(10^4)$ yr, the criteria of a quasi-constant acceleration vector during the 31 years of observational coverage is clearly given. Following Eq.~\ref{eq:minmass}, a minimum mass of $m_3$ of $2.0\pm0.4 M_{\sun}$ is however obtained for the white dwarf - much above the typical white dwarf masses of 0.5 - 0.7 $M_{\sun}$, and clearly above the Chandrasekar limit for white dwarfs of $\approx 1.4 M_{\sun}$. Assuming a mass of $0.6 M_{\sun}$ for GJ 630.1B, this object would contribute an acceleration of only $a \la 1\times10^{-8}\mathrm{m/s}^2$ on the CM Dra system. 

While GJ 630.1B can be ruled out as a source of the observed O-C variations, they may be caused by still undiscovered bodies in the  brown-dwarf mass regime (13 to 80 $M_\mathrm{Jup}$) at maximum distances of about 5 arcsec, or by a planetary-mass object (with less then 13 Jupiter masses) at a maximum distance of 2 arcsec.

\subsection{An orbiting third body\label{sec:powerspectra}}
 As shown in section~\ref{modelsel}, the sinusoidal fit matches the observed O-C times about as well as the parabolic one. From the fitted parameter, $\kappa_\phi$, we obtain with Eq.~\ref{eq:kappaphi}  and $m_3 << m_b$ a period of 
\begin{equation}
P_3=\frac{P' 2 \pi}{\kappa_\phi}=18.5\pm4.5 \mathrm{yr}
\label{eq:P3}
\end{equation}
Eq.~\ref{eq:sinampli}, with $D/c = \kappa_d$ can be rewritten as: 
\begin{eqnarray}
m_3 \sin i&=&\kappa_d \left( \frac{m_b}{M_{\sun}} \left/ \frac{P_3}{yr}\right. \right)^{2/3} 2.1 M_{\mathrm{Jup}}\ . \\
\end{eqnarray}
With the above value for $P_3$ and the fitted one for $\kappa_d$, this leads to:
\begin{equation}
 m_3 \sin i = 1.5\pm0.5M_{\mathrm{Jup}}\ .
\end{equation}
We note that such an object would have an orbital half-axis of about 5.3AU, with a maximum separation from CM Dra of about 0.35 arcsec.

\begin{figure}
   \centering
  \includegraphics[width=9.0cm]{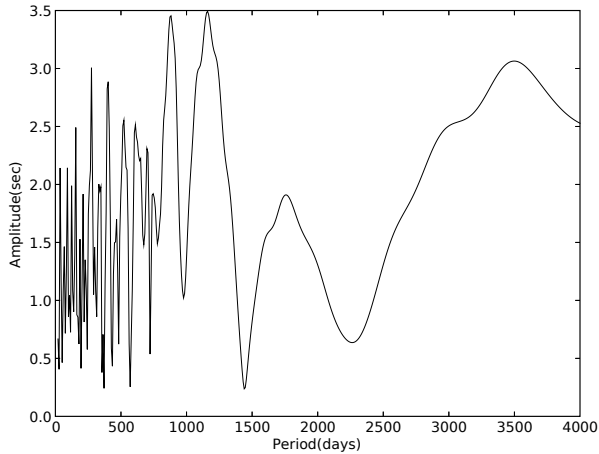}
   \caption{Power-spectrum obtained from residuals of O-C values against the parabola fit}. 
              \label{fig:powspec}
    \end{figure}

While the sinusoidal fit indicates a periodicity on a time-scale of  20 years, we also performed a search for higher frequencies. For this, the same sine-wave fitting algorithm used in DDK00 was employed. This analysis was performed on the O-C residuals using the parabola fit, and included only the relatively dense surveying that started in 1994, thereby excluding the two isolated early values obtained from \citet{lacy77}. The resulting power-spectrum is shown in Fig.~\ref{fig:powspec}.

The single broad peak at a period of 950 days and with an amplitude of 3 seconds that was apparent in \citet{ddk+98}'s Fig 2.c has now disappeared, and been replaced by several peaks with amplitudes close to 3.5  seconds. We note that this amplitude is close to the  sample standard deviation  of the yearly averaged data of $s=3.2$s. Since no single peak is outstanding, no indications for any periodicites on timescales of $\la$10 years remain.

\subsection{Applegate's mechanism}
We also evaluated the mechanism introduced by  \cite{appleypatt87} and \cite{applegate92}, which may give rise to orbital period modulations of binaries from the periodic variation of the shape - and of the quadrupole moment - of a magnetically active binary component across its activity cycles. The notion that a component of CM Dra may be magnetically active cannot be completely discarded, since several large flare events have been reported in the literature \citep{lacy77,kim+97,ddk+98}.

\begin{figure}
   \centering
  \includegraphics[width=9.0cm]{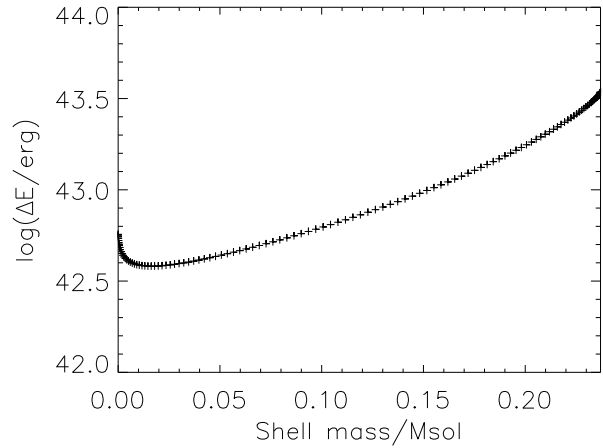}
   \caption{Energy required to vary the internal rotation of CM Dra A in order to reproduce the observed period change with Applegate's model, for all possible values of CM Dra A's shell mass. The lowest amount of energy required is $3.8\times10^{42}$ erg at a shell-mass of 0.016$M_{\sun}$}. 
              \label{fig:applegate}
    \end{figure}

In Applegate's model, the angular momentum of the entire system remains constant, but the distribution of angular momentum between the components' mutual orbit and the active star's internal rotation varies.  Energy taken up by the variation in the internal rotation has to be reflected in the active star through a corresponding luminosity variation. Applegate's original calculation considers the energy taken up by differential rotation between an inner stellar core and an outer shell of 0.1$M_{\sun}$, something which is inappropriate for either component of CM Dra, with masses of 0.207 and 0.237 $M_{\sun}$  \citep{koz+04}, respectively. We followed therefore the more general calculation introduced by \cite{brinkworth+06}, which leads to the rotational energy that has to be provided in order to explain a given period change, while doing this for any distributions of the stellar mass into core and shell. The period variation under consideration is given by \citep{applegate92}:
\begin{equation}
\Delta P = P\ 2\pi \frac{\kappa_d}{P_3}
\end{equation}
where $\kappa_d$ is the amplitude of the O-C variation, $P$ the binary period and $P_3$ the modulation period. With corresponding values taken from Table~\ref{tab:fitres} and Eq.~\ref{eq:P3}, we obtain $\Delta P = 0.010$ seconds. The {\it minimum} energy to produce such a period change for any distribution of stellar mass into core and shell (assuming a mass distribution following the Lane-Emden equation for a polytrope of n=1.5) amounts to $3.8\times10^{42}$ erg if CM Dra A is considered as the active star (see Fig.~\ref{fig:applegate}); it would be slightly higher ($4.4\times10^{42}$ erg) in the case of CM Dra B. This energy may be compared to the luminosity of CM Dra of $1.06\times10^{-2}L_{\sun}$ \citep{koz+04}, which corresponds to a release of radiant energy of $2.4\times10^{40}$ erg over the same span of $P_3=18.5$ years. With the energy required for the period change being two orders of magnitude larger than the radiant energy, Applegate's mechanism can definitively be discarded as a source of the period variations.

\section{Conclusions\label{sec:conclusion}}
The O-C timing shows a clear indication of a non-linear trend. After a review of potential causes (previous section), the remaining explanations are given by the presence of a third body. The nearly equal statistical weight of the parabolic and the sinusoidal fits currently prevents setting a clear preference for the one or the other model. For the period of the third body there exist two distinct possibilities: A Jupiter-type planet with $M \sin i$ of 1-2 $M_\mathrm{jup}$ with a period of $18.5\pm4.5$ yr, or an object such as a giant planet or heavier, with a period of hundreds to thousands of years. Intermediate-length periods of about 25 - 100 years are less likely due to poor compatibility with either the sinusoidal or the polynomial fits. Regarding shorter periodic bodies, the power-spectrum shown in Section~\ref{sec:powerspectra} excludes O-C modulations with amplitudes of larger than $\approx 3.5$ seconds with periods of $\la$10 years. The sensitivity of O-C timing detections against orbiting third bodies decreases with their period, and hence Jupiter-mass objects on such shorter periods cannot be excluded. 
A low-significance candidate for such an object with a period of about 900 days was presented in DDK00. That candidate was based on a single peak in a power-spectrum with an amplitude of 3 seconds, whereas the newer data show several peaks of amplitudes up to 3.5 seconds. The newer data therefore cannot refute that candidate, however it has become most likely to have been the result of a fortuitous combination of O-C timing values.

For long-period bodies at a quasi-constant distance and position during the observed time-span, Eq.~\ref{eq:minmass} allows us to set a maximum lateral separation from CM Dra for any given third-body mass. We also assume that any nearby body larger than about $\approx 0.1M_{\sun}$ would have become apparent in existing images of the CM Dra field, for which a large collection of CCD images exist from the TEP project \citep{ddk+98, ddk+2000}, or in 2Mass images in the IR. A maximum lateral distance of 6 arcsec, or 95AU may therefore be set for the presence of an as yet undiscovered third body of $\la  0.1M_{\sun}$. The corresponding maximum distances for undiscovered brown dwarfs or planets that could explain CM Dra's O-C timing behavior are 5 and 2 arcsec, respectively.
While the setting of third-body minimum masses (and implied brightnesses) for a given lateral distance aids in the definition and interpretation of observing projects, we note however that the observed acceleration term may be caused by relatively small third-body masses. If we take 100 years as the shortest circular period that may mimic the quasi-stationary case, such an object's orbital distance would be 16.4 AU. If it is aligned such that $|\mathbf{r}| \approx r_{\parallel}$ (e.g. $i$ close to $90\deg$), then solving Eq.~\ref{eq:accel} indicates a mass of about $1.5 M_\mathrm{jup}$. This mass may be considered the absolute minimum mass for a third body at a quasi-stationary distance, with objects at larger distances requiring larger masses.  

In conclusion, two possibilities for the source of CM Dra's timing variations remain valid: A mass of a few Jupiters on a two decade-long orbit, or an object on a century-to-millenium long orbit, with masses between 1.5 Jupiters and that of a very low mass star. Continued observations of the timing of CM Dra's eclipses over the next 5 - 10 years should, however, be decisive regarding the continued viability of the sinusoidal-fit model, and hence, about the validity of a jovian-type planet in a circumbinary orbit around the CM Dra system.

\begin{acknowledgements}
Some of the observations published in this article were made with the IAC80 telescope operated by the Instituto de Astrof\'\i sica de Tenerife in the Observatorio del Teide, and with the INT telelescope operated by the Isaac Newtown Group of Telescopes in the Observatorio del Roque de los Muchachos. This research was supported by Grant ESP2004-03855-C03-03 of the Spanish Education and Science Ministry. Some material presented here is based on work supported by NASA under the grant NNG05GJ29G, issued through the Origins of Solar Systems Program.

\end{acknowledgements}

\bibliographystyle{aa} 
\bibliography{../../../../HJDmain}
 
\end{document}